\documentclass[a4paper,11pt]{article}
\usepackage{jcappub}
\usepackage[utf8]{inputenc}
\usepackage{amsmath}
\usepackage{amssymb}
\usepackage{graphicx}
\usepackage{xcolor}
\usepackage{ulem}
\title{DESI forecast for Dark Matter-Neutrino interactions using EFTofLSS}

\author[a,b]{Markus R. Mosbech}
\author[a,c]{, Santiago Casas}
\author[a]{, Julien Lesgourgues}
\author[d]{, Dennis Linde}
\author[e]{, Azadeh Moradinezhad Dizgah}
\author[a]{, Christian Radermacher}
\author[a,f]{, and Jannik Truong}

\affiliation[a]{Institute for Theoretical Particle Physics and Cosmology (TTK), RWTH Aachen University, D-52056 Aachen, Germany}
\affiliation[b]{Institute for Theoretical Particle Physics (TTP), Karlsruhe Institute of Technology (KIT), 76128 Karlsruhe, Germany}
\affiliation[c]{Institute of Cosmology and Gravitation, University of Portsmouth, Dennis Sciama Building, Burnaby Road, 
Portsmouth, PO1 3FX, UK}
\affiliation[d]{Department of Mathematical, Physical and Computer Sciences, University of Parma, 43124 Parma, Italy}
\affiliation[e]{Laboratoire d’Annecy de Physique Théorique (CNRS/USMB), F-74940 Annecy, France}
\affiliation[f]{Department of Physics, University of California, Berkeley, CA 94720, USA}

\emailAdd{mosbech@physik.rwth-aachen.de}

\abstract{We apply the Effective Field Theory of Large Scale Structure (EFTofLSS) to non-standard models of dark matter with suppressed small-scale structure imprinted by early-time physics, here exemplified by interacting dark matter (IDM) coupled to standard model neutrinos, and 
cross-check that the EFTofLSS has no trouble replicating the real-space halo-halo power spectrum from N-body simulations. We perform forecasts for a DESI ELG-like experiment using the redshift-space power spectrum and find that, under very conservative priors on these parameters, the EFTofLSS is not expected to yield strong constraints on dark matter interactions. However, with a better understanding of the evolution of counterterms and stochastic terms with redshift, realistic IDM models could in principle be detected using the full-shape power spectrum analysis of such a spectroscopic galaxy survey.}

\begin{document}
\begin{flushright}
        {\large \tt TTK-24-40}\\
        {\large \tt TTP24-39}
\end{flushright}	
\maketitle
\flushbottom

\section{Introduction}
\label{sec:intro}
So far, the standard $\Lambda$CDM model of cosmology has proved to be a very good description of our universe \cite{Planck:2018vyg}, with some caveats such as e.g., the $H_0$ and $\sigma_8$ tensions \cite{Perivolaropoulos:2021jda} and the recent dark energy hints from DESI~\cite{DESI:2024mwx}. However, this model relies on two components whose natures are not well understood: dark energy, which in the standard picture takes the form of a cosmological constant $\Lambda$, and dark matter (DM), which in the baseline model is assumed to be non-interacting and cold. Of course, the assumption of the cold dark matter (CDM) is not truly a physical description of the DM particle, but rather a description of its apparent macroscopic behaviour. A significant amount of effort has been devoted to proposing particle physics models that can satisfy both cosmological observations and laboratory constraints, while also being produced in suitable quantities in the early universe. \cite{Bergstrom:2000pn,Bertone:2004pz,Bertone:2016nfn,Antel:2023hkf,Marsh:2024ury}. Such models will often lead to non-zero couplings with standard model particles, which could potentially give rise to signals in our experiments, or alternatively constrain the possible parameter space through the absence of detections \cite{CAST:2017uph,Billard:2021uyg,LZ:2022lsv,XENON:2023cxc,ADMX:2023rsk,OHare:2024nmr}. Cosmological observations have also been key in constraining the possible parameter space of different types of dark matter models, setting limits on interactions with baryons~\cite{Boehm:2000gq, Chen:2002yh, Boehm:2004th,Dvorkin:2013cea, Dolgov:2013una, CyrRacine:2012fz, Prinz:1998ua,Boddy:2018kfv,Slatyer:2018aqg,Xu:2018efh,Boddy:2018wzy,Becker:2020hzj}, photons~\cite{Boehm:2000gq,  Boehm:2001hm, Boehm:2004th, Sigurdson:2004zp,  McDermott:2010pa,CyrRacine:2012fz, Dolgov:2013una, Wilkinson:2013kia, Boehm:2014vja, Schewtschenko:2014fca, Schewtschenko:2015rno, Ali-Haimoud:2015pwa,  Escudero:2015yka,  Diacoumis:2017hff, Stadler:2018jin,  Stadler:2018dsa,Lopez-Honorez:2018ipk,Becker:2020hzj}, neutrinos~\cite{Boehm:2000gq, Boehm:2004th, Mangano:2006mp, Serra:2009uu, Wilkinson:2014ksa, Ali-Haimoud:2015pwa, DiValentino:2017oaw, Stadler:2019dii, Mosbech:2020ahp,Hooper:2021rjc,Akita:2023yga,Giare:2023qqn}, dark radiation species~\cite{Kaplan:2009de, Das:2010ts, Diamanti:2012tg,  Buen-Abad:2015ova, Lesgourgues:2015wza, Das:2017nub, Ko:2017uyb, Escudero:2018thh,Archidiacono:2019wdp,Becker:2020hzj,Plombat:2024kla}, DM self-interactions~\cite{Carlson:1992fn, deLaix:1995vi, Spergel:1999mh, Dave:2000ar, Creasey:2016jaq, Rocha:2012jg, Kim:2016ujt, Huo:2017vef, Markevitch:2003at, Randall:2007ph, Boehm:2000gq, Boehm:2004th}, as well as warm DM models~\cite{Dodelson:1993je,Bode:2000gq,Hansen:2001zv,Asaka:2005an,Viel:2005qj,Boyarsky:2009ix,Viel:2013fqw,Abazajian:2001nj,Boyarsky:2008xj,Dolgov:2000ew}, and axion DM models~\cite{Hui:2021tkt,Armengaud:2017nkf,Irsic:2017yje,Lidz:2018fqo,Rogers:2020ltq}.

The high-precision data expected from upcoming Stage IV galaxy surveys, including DESI~\cite{DESI:2016fyo} and Euclid~\cite{Amendola:2016saw, Euclid:2024yrr}, are poised to provide critical constraints on the unknown nature of the Universe's dark components, potentially offering unprecedented insights into the properties of dark matter and dark energy. To fully utilise the data coming from these experiments, it is necessary to make accurate predictions beyond just the linear level. The traditional method of computing non-linear predictions is via cosmological N-body simulations~\cite{Angulo:2021kes}, however, such simulations are expensive, requiring upwards of thousands of CPU hours per realisation of the cosmological parameters, and thus not suitable for e.g.,  Markov-Chain-Monte-Carlo (MCMC) parameter inference. One way to get around this is by using emulators to interpolate in a grid of simulations~\cite{Heitmann:2013bra,Heitmann:2015xma,Euclid:2018mlb,Moran:2022iwe}, but this still requires the relatively expensive computation of a grid of simulations, and a new emulator needs to be trained for each model to be tested. Another traditional method is via halo model based calculations \cite{Peacock:1996ci,Cooray:2002dia,Smith:2002dz,Takahashi:2012em,Mead:2015yca,Mead:2020vgs}, which are calibrated using N-body simulations and provide very efficient computation of the non-linear matter power spectrum. However, in general these are tuned for $\Lambda$CDM cosmologies (sometimes including neutrino masses), meaning that they are not necessarily applicable for e.g., IDM models without serious modifications. An approach developed more recently consists in modelling the full observable galaxy power spectrum using the effective field theory of large scale structure (EFTofLSS), an extension of standard perturbation theory consistently taking into account the back-reaction of small scales onto observable scales, as well as galaxy-DM biasing, redshift space distortions, and the impact of large bulk velocities~\cite{McDonald:2006mx, Baumann:2010tm, Carrasco:2012cv, Senatore:2014eva,Senatore:2014via,Senatore:2014vja, Vlah:2015sea, Blas:2015qsi, Ivanov:2018gjr, Eggemeier:2018qae, Cabass:2022avo,Linde:2024uzr}.

In this paper, we employ the EFTofLSS power spectrum predictions to forecast the sensitivity of upcoming Stage IV spectroscopic surveys, using DESI-like specifications, in constraining the dark matter scattering rate. We restrict our analysis to DM-neutrino scattering, but it is expected that our results can be generalised to other models showing similar suppressed matter power spectra, as long as the late-time growth function remains scale independent \cite{Moretti:2023drg, Tsedrik:2022cri}. For the remainder of this work, the abbreviation IDM will therefore only refer to DM-neutrino scattering, unless otherwise noted.

This paper is structured as follows: in section~\ref{sec:DMnu}, we provide a short description of our DM-neutrino scattering model, in section~\ref{sec:EFTofLSS} we provide a short description of the EFTofLSS and our implementation (for more detail see ref.~\cite{Linde:2024uzr}), in section~\ref{sec:forecast} we provide our forecast for DESI, and in section~\ref{sec:conclusion} we draw our conclusions. In appendix~\ref{sec:validation} we present a comparison of the EFTofLSS against simulations of the IDM model focusing on real-space halo power spectrum.

\section{Dark matter-neutrino interactions}
\label{sec:DMnu}
\begin{figure}
    \centering
    \includegraphics[width=0.6\linewidth]{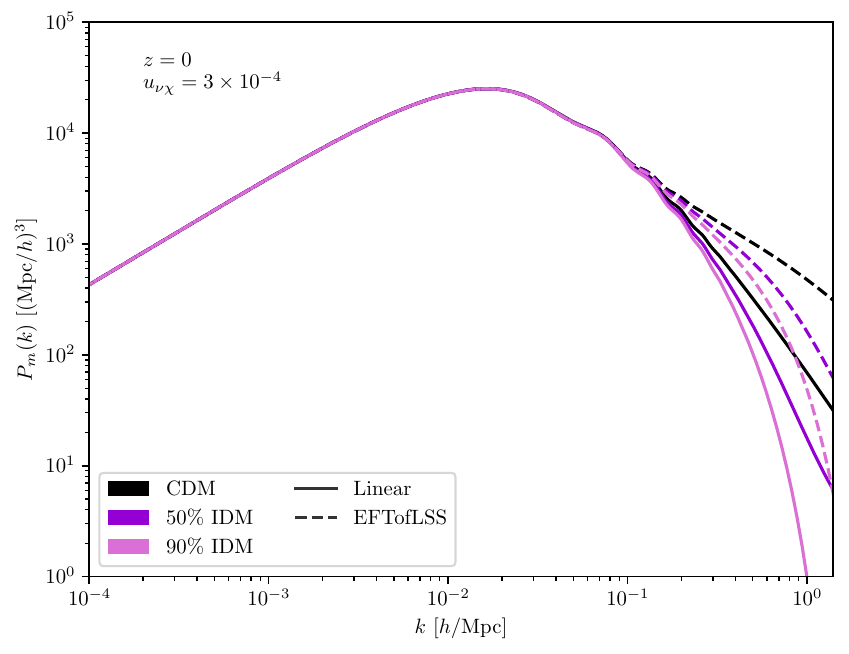}
    \caption{Linear and EFTofLSS matter power spectrum at redshift $z=0$ for two IDM scenarios, with 50\% and 90\% IDM, respectively, as well as standard CDM for reference. For the nonlinear power spectrum we have set counterterms to zero. It is evident that the scenario with the larger IDM fraction is more suppressed, but for both models the suppression starts at the same scale, as they have the same interaction strength $u_{\nu\chi}$.}
    \label{fig:pklin}
\end{figure}

The scattering of DM off neutrinos would lead to a suppression of the primordial density fluctuations below a “collisional damping scale” \cite{Boehm:2004th} and affect both the CMB and matter powers spectrum. If only part of the DM is interacting, this effect is smaller, as the non-interacting CDM can collapse as usual and form gravity wells for the IDM to fall into after its decoupling. This can be seen in figure~\ref{fig:pklin}, which shows the linear power spectrum of two different IDM scenarios, contrasted with standard CDM. On even smaller scales (relative to the suppression scale), such models will generally exhibit dark acoustic oscillations, produced from the effective pressure in the coupled fluid, analogously to the baryon acoustic oscillations imprinted from the tightly coupled baryon-photon fluid prior to recombination. Similar suppression of small-scale power from collisional damping is expected for any model with interactions between DM particles and relativistic species in the early universe\cite{Murgia:2017lwo}, such as photons \cite{Boehm:2000gq,Ali-Haimoud:2015pwa,Stadler:2018jin,Becker:2020hzj}, or dark radiation \cite{Ackerman:2008kmp,Cyr-Racine:2015ihg,Archidiacono:2019wdp,Becker:2020hzj,Plombat:2024kla}.

Limits on the scattering rate between DM and neutrinos have been  derived previously from fits to CMB data \cite{Wilkinson:2014ksa,DiValentino:2017oaw}. These limits could be improved with new data on the matter power spectrum, provided that the effect is correctly modelled on non-linear scales.

In this work we assume a scattering interaction between dark matter and standard model neutrinos, as described in Refs. \cite{Wilkinson:2014ksa,DiValentino:2017oaw,Stadler:2019dii}. A coupling between neutrinos and dark matter has been suggested as a possible source of the non-zero neutrino masses, and such couplings are exceedingly difficult to probe in laboratory experiments because of the weak coupling between neutrinos and other standard model particles \cite{Boehm:2003hm,Boehm:2006mi,Olivares-DelCampo:2017feq}. We use the approximation of massless neutrinos for the sake of simplicity in the EFTofLSS calculations, but this is unlikely to have a large impact on the final constraints, as the inclusion of neutrino masses has been shown to only have minor effects on the bounds of the interaction strength \cite{Mosbech:2020ahp,Giare:2023qqn}.

We describe the interaction strength using the effective parameter introduced in \cite{Boehm:2001hm},
\begin{equation}
    u_{\nu\chi}\equiv\frac{\sigma_{\nu \chi}}{\sigma_\mathrm{Th}}\left(\frac{m_\chi}{100\,\mathrm{GeV}}\right)^{-1},
\end{equation}
where $\sigma_\mathrm{Th}\approx6.65\times10^{-29}\,\mathrm{m}^2$ is the Thomson scattering cross-section, $m_\chi$ is the dark matter particle mass, and $\sigma_{\nu\chi}$ is the elastic scattering cross section between interacting DM particles and neutrinos. As it describes a scattering probability, it will always be positive or zero. The mass factor is included in the definition of the interaction strength to avoid a parameter degeneracy between the scattering cross section and the DM particle mass, as the interaction rate is proportional to the scattering cross section and the DM particle number density. 

The updated Euler equations for dark matter and neutrinos can be derived from the matrix element of the coupling following the approach of ref.~\cite{Cyr-Racine:2015ihg}, and are then given as \cite{Boehm:2001hm,Mangano:2006mp,Serra:2009uu,Wilkinson:2014ksa,DiValentino:2017oaw,Stadler:2019dii}
\begin{subequations}
\begin{align}
    \dot{\theta}_\nu &= k^2\psi + k^2 \left(\frac{1}{4}\delta_\nu - \sigma_\nu\right) - \dot{\mu}\left(\theta_\nu - \theta_\chi\right),\\
    \dot{\theta}_\chi &= k^2 \psi - \mathcal{H}\theta_\chi - \frac{4\rho_\nu}{3\rho_\chi}\dot{\mu}\left(\theta_\chi - \theta_\nu\right),
\end{align}
\end{subequations}
where $\theta_\nu$ and $\theta_\chi$ are the neutrino and interacting DM velocity divergences, $k$ is the comoving wave number, $\psi$ is the Newtonian gauge gravitational potential, $\mathcal{H}$ is the conformal Hubble parameter, $\rho_\nu$ and $\rho_\chi$ are the neutrino and interacting DM energy densities, and $\delta_\nu$ and $\sigma_\nu$ are the neutrino density perturbation and anisotropic stress, respectively. The interaction rate $\dot{\mu}$ is given through,
\begin{equation}
    \dot{\mu}\equiv a \sigma_{\nu \chi} n_\chi = \frac{\sigma_\mathrm{Th}}{100 \, \mathrm{GeV}} a u_{\nu\chi} \rho_\chi,
\end{equation}
where $n_\chi$ is the interacting DM particle number density.

We restrict our analysis to the simplest form of interaction, using a temperature independent cross section. In terms of the underlying particle physics model, this matches a subset of the simplified models presented in ref.~\cite{Olivares-DelCampo:2017feq}, namely models with a scalar DM and a Fermionic mediator, or vector DM with a Dirac mediator, where the DM and mediator particles are mass degenerate. There are no specific assumptions on the mass of the DM particle other than it has to be heavy enough that it can be approximated to be at rest for the interaction, i.e. it is non-relativistic and has a mass much larger than the neutrino energy.

We implement the interaction equations of refs. \cite{Stadler:2019dii,Mosbech:2020ahp} in a modified version of {\tt CLASS} \cite{Blas:2011rf} using the IDM implementation of ref.~\cite{Becker:2020hzj}.\footnote{A public version of {\tt CLASS} with these interactions implemented is currently available at \url{https://github.com/MarkMos/CLASS_nu-DM}. They will also be included in a future release on the main {\tt CLASS} repository. For the purposes of this analysis, we use the massless neutrino approximation, but both cases will be available in the code.}

\section{Effective Field Theory of Large Scale Structure}
\label{sec:EFTofLSS}

The EFTofLSS \cite{McDonald:2006mx, Baumann:2010tm, Carrasco:2012cv, Senatore:2014eva,Senatore:2014via,Senatore:2014vja, Vlah:2015sea, Blas:2015qsi, Ivanov:2018gjr, Eggemeier:2018qae, Cabass:2022avo} is an extension of standard perturbation theory that systematically incorporates three sources of nonlinearity in biased tracers of dark matter: nonlinearity in the dark matter distribution due to gravitational clustering, nonlinearity in the tracer-dark matter biasing relation, and nonlinearity arising from the peculiar velocities of tracers, which leads to redshift-space distortions. Additionally, the EFTofLSS addresses the impact of small-scale density and velocity fluctuations—which cannot be modeled perturbatively—on the overdensities of dark matter and biased tracers at large scales. It also consistently accounts for large-scale bulk flows that induce significant displacements in the matter density field, thereby affecting baryon acoustic oscillations, through the process of infrared resummation. By combining these elements, the power spectrum of biased tracers at the one-loop level includes several loop contributions, each weighted by different products of bias parameters, along with a set of counterterms to account for the impact of small-scale nonlinearities and additional terms to model stochasticity in the tracer-dark matter relation. Although these parameters could theoretically be predicted using highly accurate simulations of galaxies, current practice treats them as nuisance parameters that are marginalized over.

Over the last few years, the EFTofLSS predictions (in Lagrangian and Eulerian space) have been implemented in several publicly released codes, including {\tt CLASS-PT} \cite{Chudaykin:2020aoj}, {\tt PyBird} \cite{DAmico:2020kxu}, and {\tt Velocileptors} \cite{Chen:2020fxs,Chen:2020zjt}. In this paper, we use a new module of the {\tt CLASS}\footnote{\url{https://github.com/lesgourg/class_public}} Boltzmann code~\cite{Lesgourgues:2011re,Blas:2011rf}, {\tt CLASS-OneLoop}, which is described in ref.~\cite{Linde:2024uzr} and is planned to be released soon.

For a tractable and efficient numerical implementation of the EFTofLSS, most of the recent analytic papers and codes rely on the assumption that, in Fourier space, linear cosmological perturbations (matter density/velocity and metric fluctuations) have a separable time and wavenumber dependence. This assumption is exact in the minimal $\Lambda$CDM model with negligible neutrino masses, and still remains very good when small neutrino masses are taken into account. For non-minimal DM models that would imply a strongly scale-dependent linear growth rate, the accuracy of existing EFTofLSS codes is not a given, and their implementation would need to be scrutinized.
Fortunately, in our case and for realistic values of the interaction rate $u_{\nu\chi}$, DM decouples from neutrinos early enough to preserve a scale-independent growth rate at all redshifts relevant for non-linear structure formation, such that EFTofLSS codes are directly applicable. 
In appendix A, we confirm that the EFTofLSS prediction can reproduce the halo-halo power spectrum extracted from N-body simulations of the IDM model.

{\tt CLASS-OneLoop} features default numerical precision settings that are crucial to obtain well-converged loop integrals or a well-behaved log-Fourier transformation of the input linear power spectrum. These settings may fail for input linear spectra very different from that of the vanilla $\Lambda$CDM model, for instance due to unconverged integrals. In this work, we limit ourselves to the study of models with at least 10\% of CDM (that is, $f_\mathrm{IDM}\leq0.9$). Within this range, we have confirmed that the {\tt CLASS-OneLoop} results are always converged numerically.

For the definition of EFTofLSS parameters, we rely on the {\tt CLASS-OneLoop} parameterization, which is based on the convention in \cite{Chen:2020fxs} and detailed in \cite{Linde:2024uzr}. When modelling the one-loop galaxy power spectrum in real space at a given redshift, we have four bias parameters $(b_1, b_2, b_{{\cal G}_2}, b_{\Gamma_3})$, one counter-term $c_0$, and two stochastic parameters $(s_0, s_1)$ . 
For the one-loop galaxy power spectrum wedges in redshift space, we have the same four bias parameters, four counter-terms $(c_0, c_1, c_2, c_3)$, and four stochastic terms $(s_0, s_1, s_2, s_3)$.\footnote{Ref. \cite{Linde:2024uzr} showed that $s_1$ is mostly degenerate with other parameters. Thus, in the following, we can limit ourselves to varying the three parameters $(s_0, s_2, s_3)$. Note that $s_3$ is often refereed to as $s_4$ in other works.}

There is an on-going debate on the role of priors on nuisance parameters in the EFTofLSS. Some works have raised the issue that, when fitting the $\Lambda$CDM model to current LSS data, the posterior of some cosmological parameters tends to be prior-dominated \cite{Simon:2022lde,Holm:2023laa, Carrilho_2023}. However, given the increased sensitivity of Stage-IV surveys, this should no longer be the case in the near future \cite{Simon:2022lde,Holm:2023laa}. However, for extended cosmologies, this remains to be checked case-by-case: in models where a given physical effect (for instance, the effect of a DM interaction) would be degenerate with one (or more) EFTofLSS parameter(s), the theory would not be predictive enough in the context of this model (unless  informative priors on these EFTofLSS parameter can be obtained from simulations).

In sensitivity forecasts based on the EFTofLSS, the most optimistic assumption is that EFTofLSS parameters will be accurately measured in the future, allowing them to be fixed to specific fiducial values. In contrast, the most pessimistic assumption involves using non-informative priors and marginalizing over each parameter. Intermediate assumptions can also be made by adopting Gaussian priors on these parameters or by reducing their number through physically motivated relations among them. In the future, we expect that such priors and relations could be robustly inferred from simulations. There have been several recent works exploring the utility of simulations to obtain physically-motivated priors on EFTofLSS nuisance parameters, see e.g., \cite{Lazeyras:2017hxw,Ivanov:2024xgb,Zhang:2024thl}.

In the MCMC sensitivity forecasts of section \ref{sec:forecast}, which relies on a mock likelihood mimicking the properties of the DESI ELG sample and involving four redshift bins, we will stick to a mildly optimistic assumption concerning the four bias parameters: we will consider that their dependence on redshift is known from simulations, and thus, that each of them is known up to a free normalization parameter,
\begin{eqnarray}
b_1(z) &= \tilde{b}_1 \, b_1^{\rm fid}(z)~, \quad
b_2(z) &= \tilde{b}_2 \, b_2^{\rm fid}(z)~, \nonumber \\
b_{{\cal G}_2}(z) &= \tilde{b}_{{\cal G}_2} \, b_{{\cal G}_2}^{\rm fid}(z)~, \quad
b_{\Gamma_3}(z) &= \tilde{b}_{\Gamma_3} \,b_{\Gamma_3}^{\rm fid}(z)~,
\label{eq:bias_norm}
\end{eqnarray}
with flat uninformative priors on $(\tilde{b}_1, \tilde{b}_2, \tilde{b}_{{\cal G}2}, \tilde{b}_{\Gamma_3})$. The fiducial values of $b_{{\cal G}_2},b_{\Gamma_3}$ are chosen according to the co-evolution model while $b_2$ follows a polynomial fit; this is the same approach as used in \cite{Linde:2024uzr}. 
This co-evolution picture assumes that there were initially no biases associated with the tidal field. This implies that the late-time tidal biases ($b_{{\cal G}_2}$,$b_{\Gamma_3}$) are fully determined by local-in-matter biases ($b_1$,$b_2$) through the late-time gravitational evolution~\cite{Eggemeier:2021cam}. 
Since the interaction affects relatively small scales (comoving scales smaller than the sound horizon of the coupled fluid at DM-neutrino kinetic decoupling\footnote{The exact decoupling time, and thus suppression scale, is set by the interaction strength, with a weaker interaction decoupling earlier and thus having a suppression scale at larger $k$. See Fig.~\ref{fig:pklin} for an example.}) and decouples long before the onset of non-linear structure formation, it neither modifies the dynamics on large scales, nor induces a scale-dependency within the growth, thus we might assume the co-evolution to hold similarly to $\Lambda$CDM in our case.
For the seven counter-terms and stochastic parameters, we compare three assumptions:
\begin{itemize}
    \item In the pessimistic case, we assume seven independent parameters in each of the four redshift bins, with flat uninformative priors. We marginalize over all 28 counterterms and stochastic parameters, meaning that we have a total of 32 nuisance parameters. 
    \item In the optimistic case, like for the biases, we assume that the redshift dependence of each parameter is known (for simplicity, we consider that they are independent of $z$, but our forecast would return similar sensitivities if we assumed a fixed redshift evolution). We then marginalize over only seven counter-term and stochastic parameters, meaning that we have a total of 11 nuisance parameters. These assumptions are the same as in the DESI forecast presented in \cite{Linde:2024uzr}.
    \item In the intermediate case, we implement Gaussian priors on each counterterm (in each redshift bin) as well as on the overall normalization of the biases according to the same prescription as \cite{Simon:2022lde}. Then, we have the same nuisance parameters as in the pessimistic case, but with additional prior information. To be precise, we impose a Gaussian prior with zero mean and a standard deviation of $30\, ({\rm Mpc}/h)^2$ on each of the 28 counterterms, and a Gaussian prior with zero mean and a standard deviation of 1 for the four bias functions $b_i(\bar{z})$ evaluated at the mean redshift $\bar{z}$ of the survey. For the particular case of $b_{\Gamma_3}(\bar{z})$, the mean is shifted from 0 to $({23}/{42})[b_1(\bar{z})-1]$.
\end{itemize}
In all three cases, taking advantage of the fact that counter terms and stochastic parameters enter linearly in the expression of the galaxy power spectrum, we marginalize over them analytically, using the same method as in appendix A of \cite{Maus:2024dzi}.

\section{DESI Forecast}
\label{sec:forecast}
We limit our analysis to the ELG-like sample of DESI, which will comprise about one third of all DESI tracers. Our forecast is performed using {\tt MontePython}\footnote{\url{https://github.com/brinckmann/montepython_public}}, and employing an updated version of the DESI-ELG likelihood described in ref.~\cite{Linde:2024uzr}.\footnote{The likelihoods utilizing {\tt CLASS-OneLoop} will be publicly available on MontePython git repository upon the release of the code.} This updated version makes use of analytical marginalization~\cite{Maus:2024dzi, 2023OJAp....6E..23H} over EFT counterterms and stochastic terms. Along with several other small tweaks, this considerably speeds up the convergence of our MCMC chains.\footnote{For instance, even in the pessimistic case, we could get 6 chains to converge on a timescale of hours using 24 cores on a laptop.} The updated version of the likelihood will be released on the public MontePython repository at the same time as CLASS-OneLoop. The survey specifications for the ELG sample are shown in tab.~\ref{tab:forecast_surveyparams}, including redshift ranges, corresponding bin volumes, mean number densities, and linear biases. Although the bias parameter $b_{\Gamma_3}$ enters the galaxy power spectrum linearly, we do not perform an analytic marginalization over it in order to retain the ability to plot marginalized posteriors and contours for all bias parameters.

We perform our analysis with two different fiducial cosmologies, a 100\% CDM fiducial to ascertain the total constraining power of the DESI ELG sample in the case of a universe without suppressed structure formation, as well as a 90\% IDM fiducial scenario to determine the discovery power, in case our universe would resemble such a scenario. The fiducial values of cosmological parameters and the assumed priors are listed in tab.~\ref{tab:cosmo_fid}, For each of the relative bias normalization parameter $\tilde{b}_i$ defined in \eqref{eq:bias_norm}, we use a fiducial value of 1. For the counterterms and stochastic terms we use $\{c_{0}, c_{1}, c_{2}, c_{3}\}=\{-10, 20, 20, 20\}$ and $\{s_0,s_2,s_3\}=\{0,0,0\}$, following ref. \cite{Linde:2024uzr}. 

We also examine the different levels of prior information on the biases, counterterms, and stochastic terms, presented as \textit{pessimistic}, \textit{optimistic} and \textit{intermediate} at the end of sec.~\ref{sec:EFTofLSS}. Additionally, we always use a Gaussian Big Bang Nucleosynthesis (BBN) prior on $\omega_{\rm b}$, centered on our assumed fiducial value and with a width $\sigma(\omega_{\rm b}) = 0.00036$ reflecting the current BBN sensitivity \cite{Pisanti:2020efz}.

\begin{table}[]
    \centering
    \begin{tabular}{|cc|c|c|c|}
        \hline
        $z_{\mathrm{min}}$ & $z_{\mathrm{max}}$ & $V_c(z)$ $[\mathrm{Gpc}^3 \, h^{-3}]$ & $n_g(z)$ $[10^{-4}\, \mathrm{Mpc}^{-3} \, h^3]$ & $b_1(z)$  \\ \hline
        $0.6$ & $0.85$ & $6.680$ & $5.811$ & $1.223$ \\
        $0.85$ & $1.1$ & $9.045$ & $8.010$ & $1.373$ \\
        $1.1$ & $1.35$ & $10.811$ & $4.023$ & $1.527$ \\
        $1.35$ & $1.7$ & $17.081$ & $0.915$ & $1.716$ \\ \hline
    \end{tabular}\vspace{0.15in}
    \caption{Parameters for a DESI ELG-like 4 bin survey, redshift bin edges $z_\mathrm{min}$ and $z_\mathrm{max}$, comoving bin volume $V_c(z)$, galaxy number density $n_g(z)$, and linear bias $b_1(z)$. The likelihood is described in detail in ref.~\cite{Linde:2024uzr}. } \vspace{-0.15in}
    \label{tab:forecast_surveyparams}
\end{table}

\begin{table}
    \centering
    \begin{tabular}{|c|ccc|c|}
    \hline
    Parameter     &  & Fid. Value & & Prior  \\
    \hline
         $\omega_{\rm b}$ & & 0.0224 & & $\mathcal{N}\left(0.022445 , 0.00036\right)$ \\
         $\omega_{\rm m}$ &  & 0.1424 & & Flat, unbounded \\
         $h$ &  & 0.6736 & & Flat, unbounded \\
         $n_{\rm s}$ &  & 0.9649 & & Flat, unbounded \\
         $\ln 10^{10} A_{\rm s} $ &  & 3.0448 & & Flat, unbounded \\
         \hline
         & CDM & & IDM90 & \\
         \hline
         $f_\mathrm{idm}$& 0 &  & 0.9 & Not varied \\
         $u_{\nu\chi}$& / & & $5\times10^{-5}$ & Flat, $u_{\nu\chi}\geq 0$ \\
         \hline
    \end{tabular}
    \caption{Cosmological parameters for each fiducial model and the priors used in the DESI ELG forecast.}
    \label{tab:cosmo_fid}
\end{table}

\subsection{Optimistic results}

\begin{figure}[t]
    \centering
    \includegraphics[width=\linewidth]{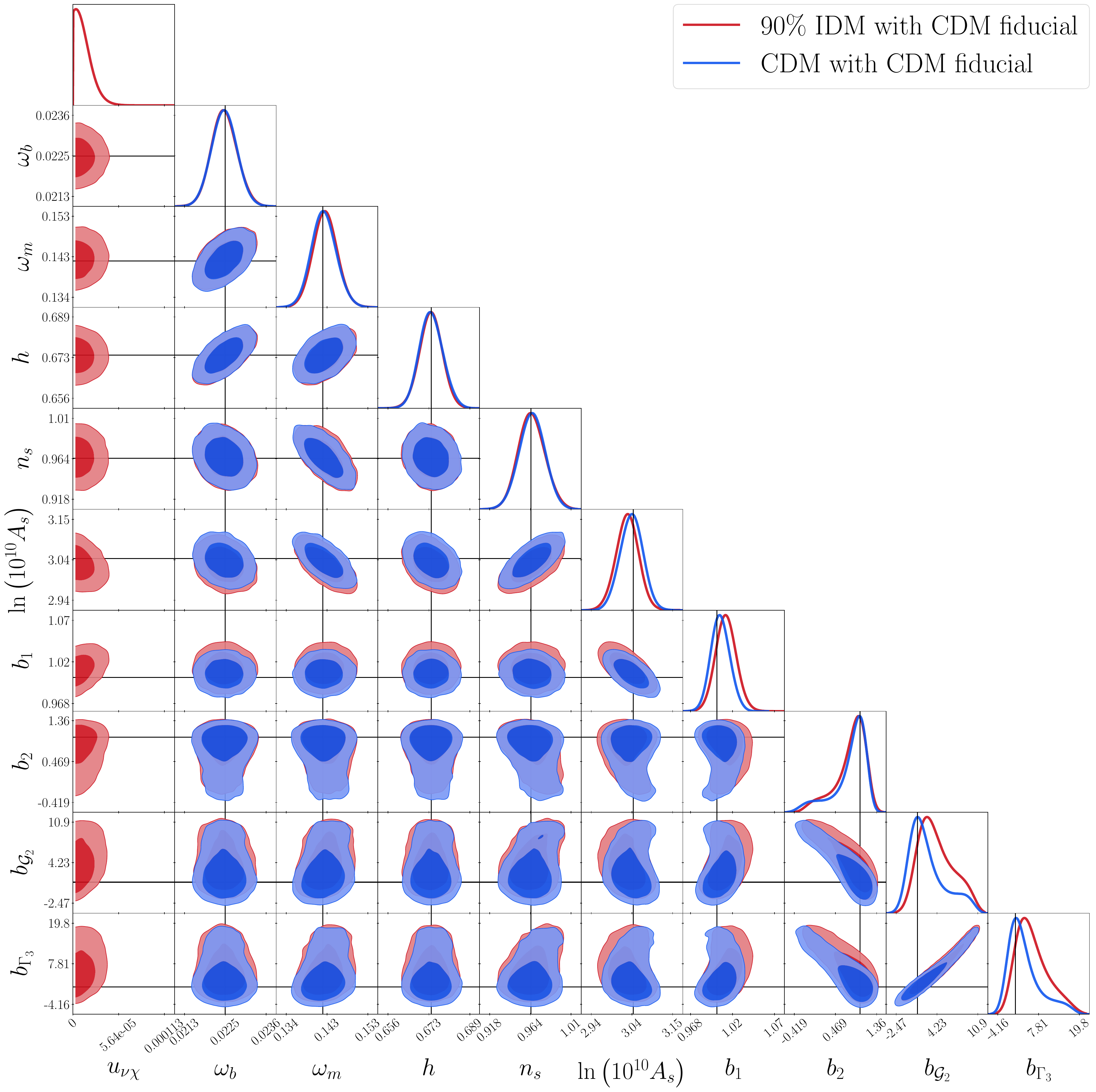}
    \caption{Triangle plot of the one-dimensional posteriors and two-dimensional (68\% and 95\%) iso-credibility contours of the MCMC chains using a CDM-only fiducial and constant counterterms, with fiducial values marked with black lines. It is evident that the fiducial parameter values are well recovered using both the CDM-only (as in fiducial) and 90\% IDM models, indicating that such an experiment can yield robust limits on the DM-neutrino interaction strength.}
    \label{fig:triangle_cdmfid}
\end{figure}

\begin{figure}
    \centering
    \includegraphics[width=\linewidth]{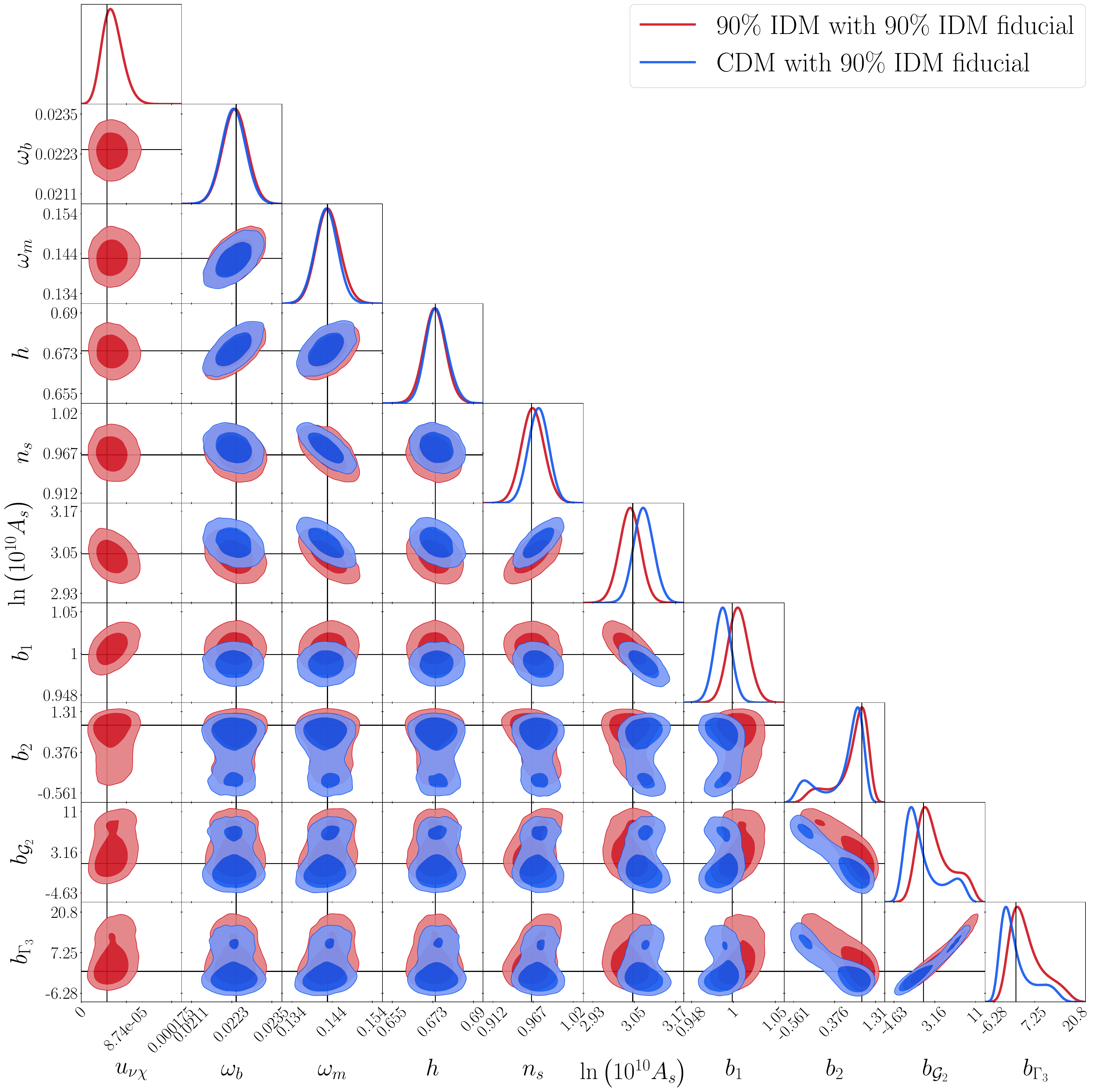}
    \caption{Triangle plot of the ne-dimensional posteriors and two-dimensional (68\% and 95\%) iso-credibility contours of our MCMC chains using a 90\% IDM fiducial, with fiducial values marked with black lines, including only those parameters with significant difference between the 90\% IDM and CDM chains. It is evident that the fiducial parameter values are well recovered with the 90\% IDM models, while there are $\sim 1\sigma$ shifts in central values recovered in the CDM only chains.}
    \label{fig:triangle_idm90fid_shift}
\end{figure}

We begin by focusing on the case of optimistic priors on EFTofLSS parameters, corresponding to 11 free nuisance parameters with known redshift dependencies. We present our results in figure~\ref{fig:triangle_cdmfid} for a pure CDM fiducial model, and in figure~\ref{fig:triangle_idm90fid_shift} for a fiducial model with 90\% IDM. In each figure, the two fitted models assume either pure CDM or 90\% of IDM (blue and red respectively, in Figs.~\ref{fig:triangle_cdmfid} and \ref{fig:triangle_idm90fid_shift}). Note that the pure CDM model is always nested in the IDM model: it corresponds to the limit of a negligible interaction rate, $u_{\nu\chi}\rightarrow0$.

As a preliminary check, we perform an MCMC run assuming pure CDM both in the fiducial model and in the fitted model. In this case, as expected, we correctly recover all cosmological and bias parameters, with unbiased marginalized posteriors.

In a second step, we fit the same pure CDM fiducial model with the 90\% IDM model. This is useful to estimate the sensitivity of the experiment to the DM-neutrino interaction rate. This run also correctly recovers all parameters, with now a small bias (of less than 1$\sigma$) on the parameters that are most correlated with $u_{\nu\chi}$, namely the primordial amplitude $A_{\rm s}$ and some of the bias parameters. This reflects the fact that the effect of an interaction can to some extent be counteracted by shifting the amplitude and biases. The most interesting conclusion from this run is that a DESI ELG-like sample should be able to set a credible interval for the the DM-neutrino interaction strength bounded by $u_{\nu\chi}\leq3.5\times10^{-5}$ at the 95\% confidence level (C.L.), which would be a stronger limit than that from current CMB and BAO data from Planck, ACT, SPT and BOSS, see ref.~\cite{Giare:2023qqn}.

Using a fiducial cosmology with the same values for the standard cosmological parameters, but containing 90\% IDM with $u_{\nu\chi}=5\times10^{-5}$, we also find that the MCMC accurately recovers the fiducial values when letting $u_{\nu\chi}$ vary, setting a credible interval on the interaction strength of $u_{\nu\chi}=6.1^{+4.1}_{-3.9}\times10^{-5}$ (95\% C.L.), and thus showing a preference for IDM over CDM at the $\sim 3\sigma$ level. The fact that the posterior is centered on 6 rather than $5\times10^{-5}$ could be a prior effect, as a flat $u_{\nu\chi}\geq0$ prior was used. Running a CDM-only MCMC with the IDM fiducial cosmology, the fiducial cosmological and bias parameters are recovered to within $\sim1\sigma$, but with shifted central values for $A_{\rm s}$ and bias parameters in particular, see figure~\ref{fig:triangle_idm90fid_shift}. There is a slight change in $n_{\rm s}$ as well, since at large $k$, where the power spectrum is suppressed by the interaction, the changed slope could be mimicked by a change in the primordial tilt. However, since this effect is scale-dependent, a slight change in the tilt and amplitude will be a ``compromise solution'', slightly improving the fit at large $k$ without worsening the fit at small $k$ too much. This shows again that the model may try to mimick the IDM effect with a shifted value of these parameters. However, the quality of the best fit is worse\footnote{We note that the $\chi^2$ used here is provided after the marginalization over stochastic- and counterterms. We expect both scenarios to be roughly equally penalised by this. Additionally, since we are not using a physically motivated model for the value of these parameters, using the marginalized $\chi^2$ should be a conservative choice, avoiding potentially fine-tuned values.}, with $\Delta \chi^2 = -6.88$, defined as
\begin{equation}
    \Delta \chi^2 = \chi^2_{\min ,  \text{IDM}} - \chi^2_{\min ,  \text{CDM}},
\end{equation}
or an Akaike information criterion of $\Delta \mathrm{AIC} = -4.88$, defined as
\begin{equation}
    \Delta \mathrm{AIC} = \chi^2_{\min ,  \text{IDM}} - \chi^2_{\min ,  \text{CDM}} + 2(N_\text{IDM}-N_\text{CDM}),
\end{equation}
where $N_i$ is the number of model parameters of model $i$. Having 90\% of the dark matter interacting is almost equivalent to all of the dark matter interacting, so leaving this fraction fixed, we treat the IDM90 model as having only one extra free parameter relative to $\Lambda$CDM, indicating a weak preference for the interacting model \cite{Jeffreys:1939xee,Nesseris:2012cq}. However, this analysis relies on strong assumptions on the stochastic parameters and counterterms. More conservative assumptions can considerably decrease the constraining power, as described in section~\ref{subsec:pessimistic}.

\subsection{Comparison with intermediate and pessimistic cases}
\label{subsec:pessimistic}

Our \textit{optimistic} case assumes the counterterms and stochastic terms are redshift-independent parameters. If instead, we considered these parameters to have a known redshift evolution and an unknown overall amplitude, as we are doing for the bias parameters, we would obtain similar forecast errors, since the number of free parameters and the degree of uncertainty would be identical. In the future, simulations of galaxy populations with properties that match those of observed samples may provide better insight into the redshift dependence of these parameters, but probably within some remaining uncertainty. On the most conservative side, we may assume that the redshift dependence is totally unconstrained and that counterterms and stochastic terms are independent from each other in each redshift bin, which corresponds to our \textit{pessimistic case}. We expect that the pessimistic and optimistic cases bracket realistic constraints from the actual observational data. Our \textit{intermediate} case, which features additional Gaussian priors, is an example of possible intermediate assumptions, with prior widths set in a relatively arbitrary way. 

\begin{figure}
    \centering
    \includegraphics[width=\linewidth]{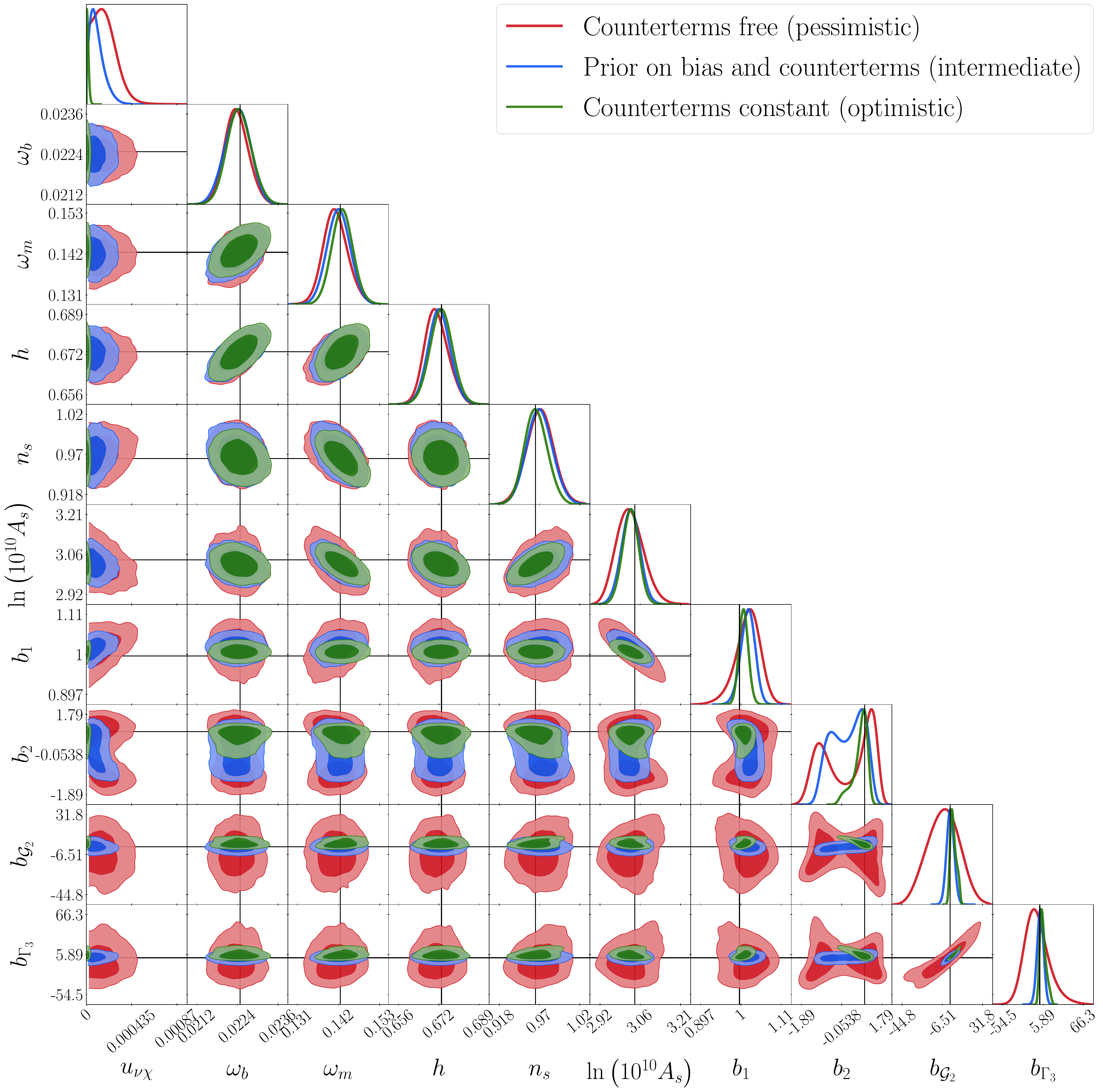}
    \caption{Triangle plot of the 2d  posteriors of the optimistic (green), intermediate (blue), and pessimistic (red) analyses with a CDM fiducial cosmology. It is clear that stronger priors on stochastic and counterterms greatly improve the constraining power for $u_{\nu\chi}$ in particular.}
    \label{fig:cdmfid-variations}
\end{figure}

\begin{figure}
    \centering
    \includegraphics[width=\linewidth]{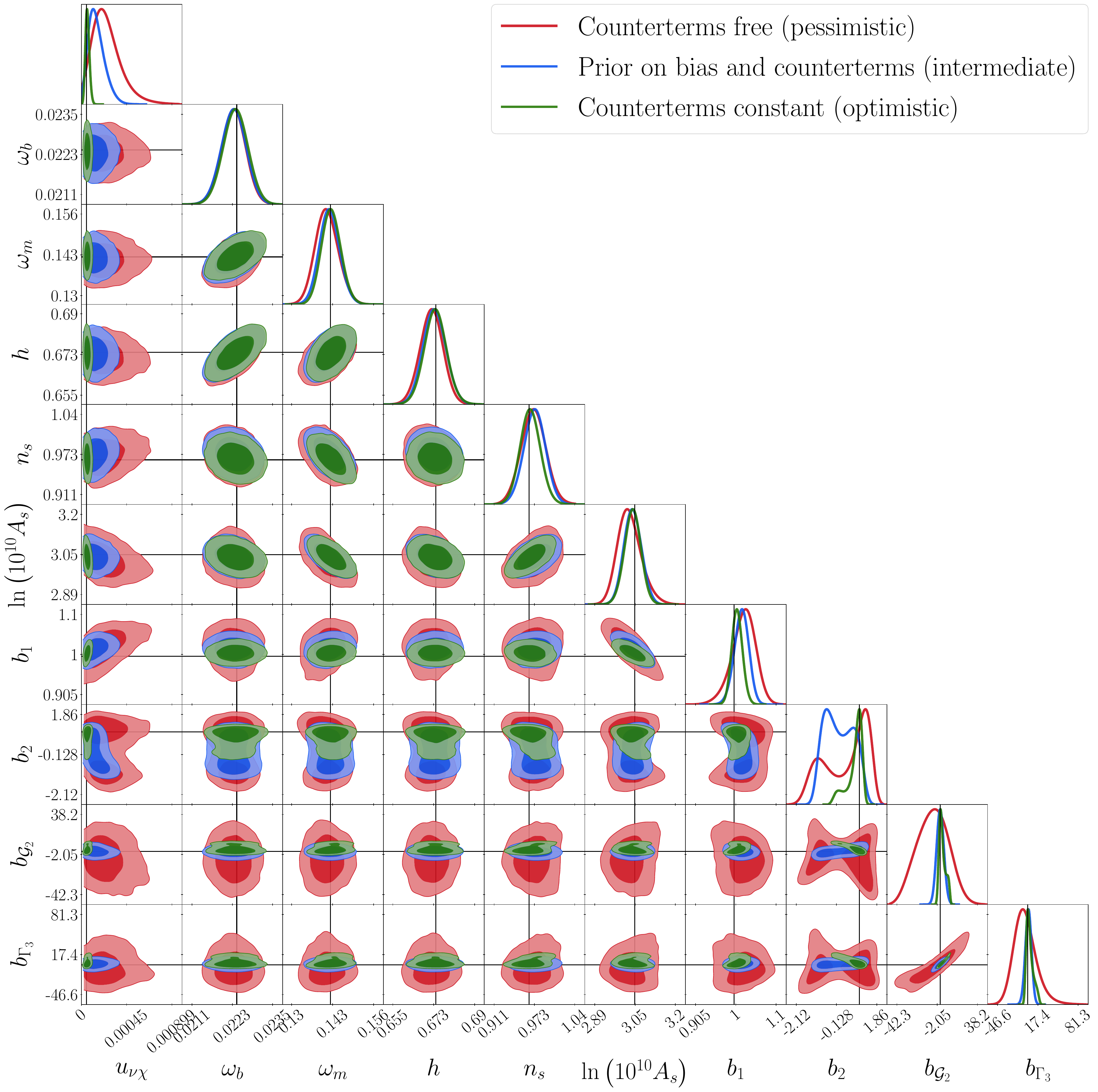}
    \caption{Triangle plot of the 2d  posteriors of the optimistic (green), intermediate (blue), and pessimistic (red) analyses with a 90\% IDM fiducial cosmology. Again, it is clear that stronger priors on stochastic and counterterms greatly improve the constraining power for $u_{\nu\chi}$ in particular.}
    \label{fig:idmfid-variations}
\end{figure}

The posteriors for the intermediate and pessimistic cases are shown in Figs.~\ref{fig:cdmfid-variations} and \ref{fig:idmfid-variations}, along with the optimistic case for reference.

The pessimistic case shows greatly reduced constraining power for the interaction strength $u_{\nu\chi}$ as well as all four bias parameters. Additionally the pessimistic case finds a significantly biased value of $u_{\nu\chi}$ for the IDM90 fiducial cosmology with the posterior centered at a value more than $1\sigma$ larger than the fiducial. However, the posteriors on the other cosmological parameters generally do not show significant change, except for a small shift in $A_{\rm s}$ (which makes sense, as this is the cosmological parameter showing greatest degeneracy with the interaction strength). With the CDM fiducial, the pessimistic case yields an upper limit for the interaction strength of $u_{\nu\chi}\leq 3.9\times 10^{-4}$ (95\% C.L.), which is the same order of magnitude as Planck CMB constraints\cite{Mosbech:2020ahp}. All upper limits and inferred values are collected in Table~\ref{tab:u_vals}. It was to be expected that the parameter most affected by the pessimistic assumption would be $u_{\nu\chi}$, since this is the only cosmological parameter in our analysis that affects the power spectrum only on small (mildly non-linear) scales, and thus, which can have a strong correlations with EFT parameters. The fact that $A_{\rm s}$ is also slightly impacted simply follows from the small correlation between this parameter and $u_{\nu\chi}$. All chains are well converged with $R-1 \lesssim 0.01$ for all parameters, except the $b_2$ bias in our pessimistic analysis with CDM prior, which has $R-1\approx0.02$.

\begin{table}[]
\centering
\begin{tabular}{|l|c|}
\hline
             & 95\% CL upper limit (CDM fid.)        \\ \hline
Optimistic   &    $<3.49\times 10^{-5}$              \\
Intermediate &    $<2.48\times 10^{-4}$              \\
Pessimistic  &    $<3.93\times 10^{-4}$              \\ \hline
             & Mean $\pm2\sigma$ (IDM90 fid.)        \\ \hline
Optimistic   &  $6.08^{+4.08}_{-5.23}\times 10^{-5}$ \\
Intermediate &  $1.54^{+1.62}_{-1.42}\times 10^{-4}$ \\
Pessimistic  &  $2.54^{+2.94}_{-2.50}\times 10^{-4}$ \\ \hline
\end{tabular}
    \caption{Forecast upper limits (for CDM fiducial) and inferred values (for IDM90 fiducial) of the interaction parameter $u_{\nu\chi}$.}
    \label{tab:u_vals}
\end{table}

In the intermediate case, the constraining power is also significantly reduced compared to the optimistic case, but not to the same extent as the pessimistic case. The posteriors of the biases are also significantly narrowed. It is not obvious if this is solely due to the imposed priors on the biases themselves, or if the priors on counterterms also play a role due to possible degeneracies. With the CDM fiducial, the upper limit on the interaction strength is reduced to $u_{\nu\chi}\leq 2.5\times 10^{-4}$, which is a minor improvement. Using the IDM90 fiducial, the result also remains biased towards stronger interactions with the posterior centered $1\sigma$ above the fiducial.. The difference in the constrains on $A_{\rm s}$ between the pessimistic and intermediate cases is driven by the better constraint on $b_1$ due to the imposed prior, since these two parameters are highly degenerate.

Overall, the impact of loosening our assumptions on stochastic terms and counterterms indicates that there is a significant degeneracy between the interaction parameter $u_{\nu\chi}$ and the stochastic contributions or counterterms. Our results in the intermediate and the pessimistic cases indicate that in the absence of well-motivated informative priors on these nuisance parameters, the limits on dark matter-neutrino interactions from galaxy clustering power spectrum using the EFTofLSS model will likely not to be competitive with those from the CMB. If the assumptions on the redshift evolution of the four bias parameters were loosened, this could lead to similar degradation in constraining power. This conclusion is expected to be extendable to other models with small-scale suppression of linear matter power spectrum. However, the constraining power observed in the optimistic case suggests that improved constraints may be achievable in the future provided that a physically-motivated model for the redshift dependence of the EFTofLSS parameters can be established.

\section{Conclusion}
\label{sec:conclusion}
In this work, we have investigated the applicability of EFTofLSS for constraining exotic dark matter scenarios beyond standard CDM, as exemplified by DM-neutrino scattering, with spectroscopic galaxy surveys. In presence of such an interaction, the shape of the matter power spectrum is altered in the early universe while the growth factor and growth rate remain scale-independent during structure formation. Thus, the EFTofLSS formalism can be employed to predict the galaxy power spectrum in redshift space, in the same way as for $\Lambda$CDM. In appendix~\ref{sec:validation}, we confirm that the EFTofLSS prediction can reproduce the real-space halo-halo power spectrum extracted from N-body simulations of IDM models.

In our main analysis, we performed a sensitivity forecast for the parameters of the IDM model, assuming that the redshift space power spectrum is measured by an ELG-like sample of the DESI survey. We use a  parametrization of the redshift evolution of the biases described in our previous work \cite{Linde:2024uzr} to reduce the freedom in the EFT parameters. By using analytic marginalization over the counterterms and stochastic terms, we can also relax some of the optimistic assumptions of ref.~\cite{Linde:2024uzr}, allowing the EFTofLSS counterterms to vary independently in each redshift bin. We perform four MCMC forecasts, using either a pure CDM or a 90\% IDM cosmology as a fiducial model and as a fitted model, to determine both constraining and discovery power. 

Our most optimistic forecast keeps the counterterms and the parameters describing stochastic contributions constant in redshift as in the analysis of ref.~\cite{Linde:2024uzr}, yielding strong constrains on the interaction strength, more constraining that the current SPT and ACT CMB constraints of ref.~\cite{Giare:2023qqn}. However, the optimistic forecast is probably unrealistic, unless future developments enable us get insight into the redshift dependence of the EFTofLSS nuisance parameters. We therefore run also a pessimistic case, letting them vary independently in each redshift bin, and an intermediate case, putting Gaussian priors on the counterterms and biases. In both the intermediate and pessimistic cases, the constraining power is significantly reduced, the uncertainties increasing by an order of magnitude and the inferred interaction strength showing significant bias. 

Our conclusion is therefore that under conservative assumptions for the least-constrained nuisance parameters, namely the counterterms and stochastic contributions, EFTofLSS is currently not able to significantly improve constraints on dark matter models with suppressed small scale structure. However, our optimistic forecast indicates that the constraining power could in principle be greatly improved with a better understanding of these parameters, as illustrated by Figs.~\ref{fig:cdmfid-variations} and \ref{fig:idmfid-variations}. The inclusion of higher-order statistics---like the bispectrum---may help in breaking degeneracies between DM parameters and stochastic parameters or counterterms, and is left for a future study.

\acknowledgments
The authors would like to thank Felix Kahlh{\"o}fer for useful discussions.
The authors gratefully acknowledge the computing time provided to them on the high-performance computer Lichtenberg at the NHR Center NHR4CES@TUDa, funded by the German Federal Ministry of Education and Research (BMBF) and the Hessian Ministry of Science and Research, Art and Culture (HMWK). JL and MM acknowledge support from the DFG grant LE 3742/8-1. Some tests of the likelihood were performed with computing resources granted by RWTH Aachen University under projects \texttt{thes1654} and \texttt{rwth1389}. SC acknowledges support by the Science and Technology Facilities Council (grant number ST/W001225/1). The authors gratefully acknowledge the computing time provided to them at the NHR Center NHR4CES at RWTH Aachen University (project number p0021792). This is funded by the Federal Ministry of Education and Research, and the state governments participating on the basis of the resolutions of the GWK for national high performance computing at universities (www.nhr-verein.de/unsere-partner).

\appendix
\section{Comparison with N-body simulations}
\label{sec:validation}

\begin{figure}[b]
    \centering
    \includegraphics[width=\linewidth]{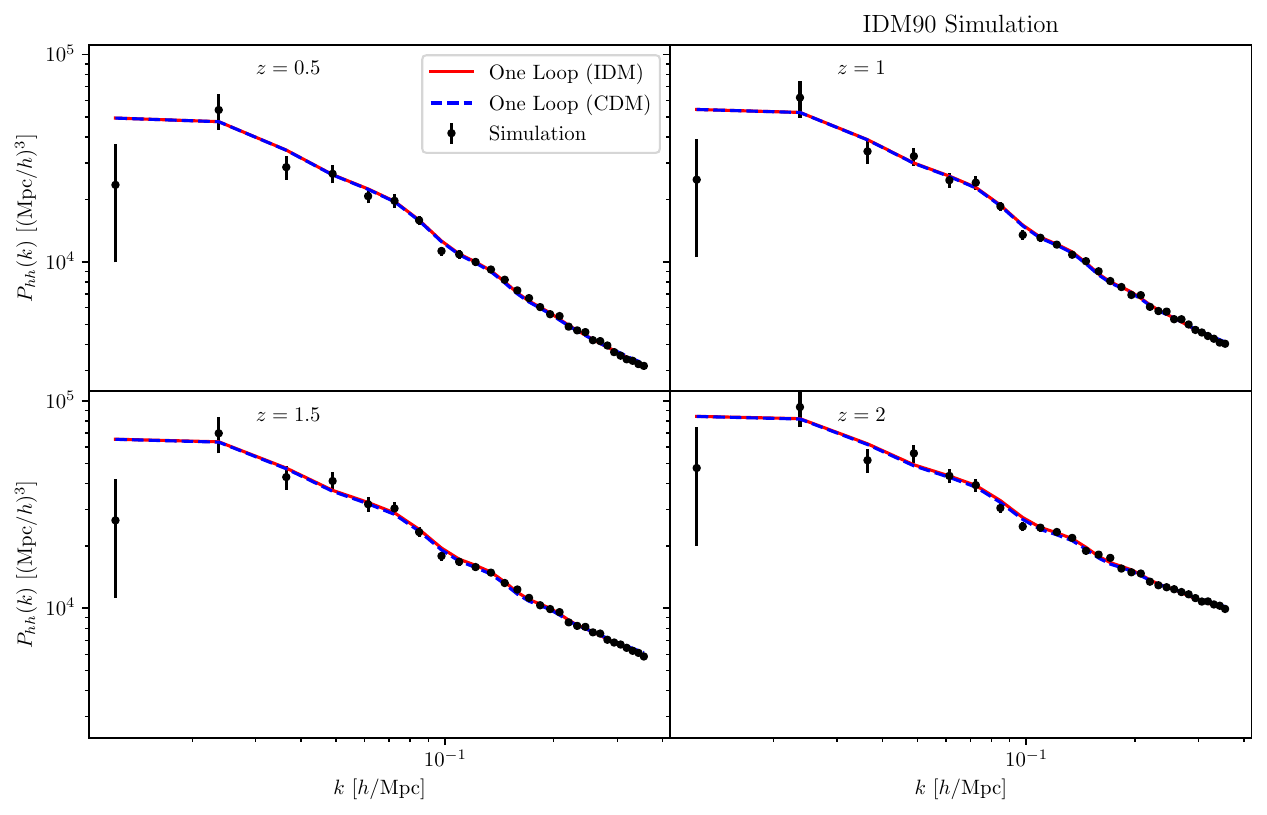}
    \caption{Halo-halo power spectrum from the 90\% IDM N-body simulation and {\tt CLASS-OneLoop} best-fit for the, showing a good agreement at all relevant redshifts. When letting all EFT parameters vary, it is clear that the CDM model can fit simulations roughly as well as IDM, enough for the models to be indistinguishable. The residuals are shown in figure~\ref{fig:Phh_residuals}.}
    \label{fig:Phh_idm90}
\end{figure}

\begin{figure}[htbp!]
    \centering
    \includegraphics[width=\linewidth]{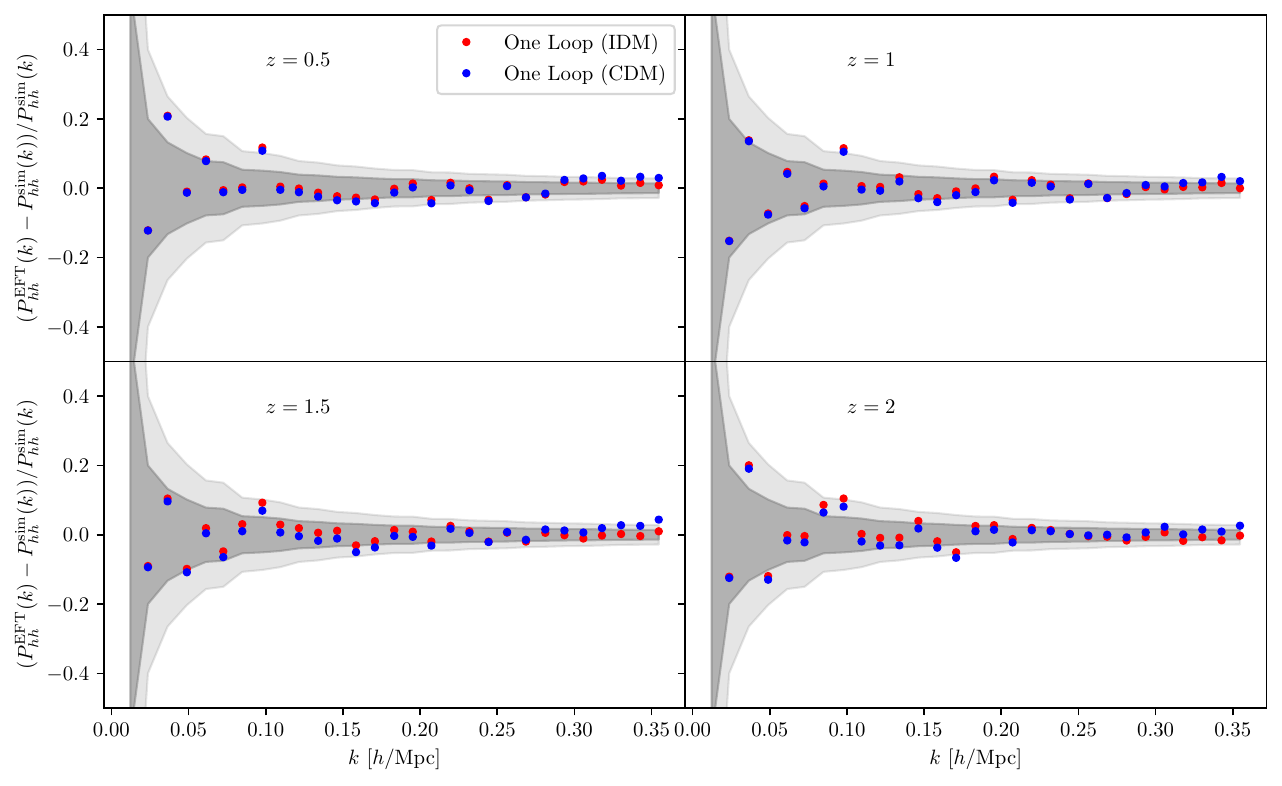}
    \caption{Residuals of the {\tt CLASS-OneLoop} best-fit results relative to the simulation. The results generally fall within the 1 and 2$\sigma$ regions, shown as dark and light grey bands, as expected for a good fit. These regions are obtained from the Gaussian covariance of the simulation.}
    \label{fig:Phh_residuals}
\end{figure}

We perform a preliminary check to ensure that EFTofLSS can reproduce simulation results for our IDM models. For this, we run a suite of N-body simulations covering the scales and redshift range of interest. We use $N=512^3$ particles and a box size of 512 Mpc$/h$, producing a simulation with 90\% IDM, and one with 50\% IDM, as well as a 100\% CDM simulation for comparison. We fix the interaction strength at $u_{\nu\chi}=5\times10^{-5}$ (roughly matching the best-fit values of \cite{Giare:2023qqn}). This value of $u_{\nu\chi}$ is likely already in tension with the galaxy luminosity function~\cite{Mosbech:2022nkk}, so we expect that being able to fit this and weaker interactions (i.e. CDM-only) means the method should be applicable across realistic values of the interaction strength.

We perform our simulations with the code {\tt Gadget-4} \cite{Springel:2020plp} using initial conditions generated from linear {\tt CLASS} spectra with the built-in {\tt ngenic} algorithm, passing as an input the power spectra computed with our modified version of {\tt CLASS}. The rest of the simulation parameters are listed in tab.~\ref{tab:sim_params}. We output snapshots at redshifts $z_\mathrm{out}=[0.5,\,1,\,1.5,\,2]$. 
\begin{table}[t]
    \centering
    \begin{tabular}{|c|c|}
    \hline
        $\omega_{\rm m}$ & 0.31\\
    \hline
        $\omega_{\rm b}$ & 0.049 \\
    \hline
        $A_{\rm s}$ & $2.11\times10^{-9}$ \\
    \hline
        $n_{\rm s}$ & 0.968 \\
    \hline
        $h$ & 0.678 \\
    \hline
        $z_\mathrm{ini}$ & 63 \\
    \hline
    $u_{\nu\chi}$ & $\{0 , \, 5\times 10^{-5} $\} \\
    \hline
    \end{tabular}
    \caption{Cosmological parameters of the two simulations. For $u_{\nu\chi}$, the value 0 (meaning 100\% non-interacting CDM) is used for the comparison simulation, while $5\times 10^{-5}$ is used as an example of an interacting scenario.}\vspace{-0.15in}
    \label{tab:sim_params}
\end{table}

We use {\tt Gadget-4}'s on-the-fly {\tt FoF} and {\tt subfind} halo finders to generate our halo catalog. From these, we use {\tt nbodykit} \cite{handBccpNbodykitNbodykit2018} to compute the halo-halo power spectrum $P_{hh}(k)$ at each redshift. 

Keeping cosmology fixed, we fit the EFTofLSS parameters to the simulation results, letting the parameters vary freely at each redshift. We find that both the IDM and CDM models are able to reproduce the 90\% IDM simulation results, to within the errorbars of the simulation, as illustrated in Figs.~\ref{fig:Phh_idm90} and \ref{fig:Phh_residuals}.

A full validation study against simulation results in redshift space and varying cosmological parameters will be presented in forthcoming work.

\bibliography{bibliography}

\end{document}